\documentclass[twocolumn,epjc3]{svjour3}
\smartqed  % flush right qed marks, e.g. at end of proof
\RequirePackage{graphicx}

\journalname{Eur. Phys. J. C}

\usepackage{dcolumn}% Align table columns on decimal point
\usepackage{bm}% bold math
\usepackage{ifpdf}
\usepackage{hyperref}
\usepackage{dcolumn}
\usepackage{bm}
\usepackage[spanish,english]{babel}
\usepackage{amsfonts}
\usepackage{amssymb}
\usepackage{graphicx}
\usepackage[latin1]{inputenc}

\newcommand{\E}{\mathcal{E}}
\newcommand{\LL}{\mathcal{L}}

\begin{document}

\title{Geonic black holes and remnants in Eddington-inspired Born-Infeld gravity}

\author{Gonzalo J. Olmo \thanksref{e1,addr1}
        \and
        D. Rubiera-Garcia  \thanksref{e2,addr2}
        \and
        Helios Sanchis-Alepuz \thanksref{e3,addr3}
        }

\thankstext{e1}{e-mail: gonzalo.olmo@csic.es}
\thankstext{e2}{e-mail: drubiera@fisica.ufpb.br}
\thankstext{e3}{e-mail: helios.sanchis-alepuz@theo.physik.uni-giessen.de}

\institute{Departamento de F\'{i}sica Te\'{o}rica and IFIC, Centro Mixto Universidad de
Valencia - CSIC. Universidad de Valencia. Burjassot-46100, Valencia, Spain \label{addr1}
        \and
        Departamento de F\'isica, Universidade Federal da
Para\'\i ba, 58051-900 Jo\~ao Pessoa, Para\'\i ba, Brazil \label{addr2}
        \and
        Institute of Theoretical Physics, Justus-Liebig University of Gie{\ss}en,
Heinrich-Buff-Ring 16, 35392, Gie{\ss}en, Germany \label{addr3}
        }

\date{Received: date / Accepted: date}

\maketitle

\begin{abstract}
We show that electrically charged solutions within the Eddington-inspired Born-Infeld theory of gravity replace the central singularity by a wormhole  supported by the electric field. As a result, the total energy associated with the electric field is finite and similar to that found in the Born-Infeld electromagnetic theory. When a certain charge-to-mass ratio is satisfied,  in the lowest part of the mass and charge spectrum the event horizon disappears  yielding stable remnants. We argue that  quantum effects in the matter sector can lower the mass of these remnants from the Planck scale down to the TeV scale.



\end{abstract}

\section{Introduction}

Historically, the taming of singularities in classical field models has driven a great deal of research. A particularly elegant example is the nonlinear extension of Maxwell electrodynamics introduced by Born and Infeld \cite{BI} to remove the divergence of both the Coulomb field and the self-energy of point particles. In this \emph{determinantal} form of the classical action, the modified field (the \emph{BIon} \cite{Gibbons}), is everywhere bounded but generated by a distributional source. This specific form of nonlinear electrodynamics arises in the low-energy limit of certain string theories \cite{strings-BI}.

Recovering the idea of the determinantal form of the gravitational action suggested by Eddington \cite{Eddington}, an Eddington-inspired Born-Infeld action (EiBI) for the gravitational field has been introduced recently \cite{Banados,Deser}. In order to avoid troubles with higher-order derivatives and ghosts, EiBI gravity is formulated in the Palatini approach, which means that metric and connection are regarded as physically independent entities \cite{olmo11}. This implies that the connection is determined by the field equations, not constrained {\it a priori} to any particular form.

The EiBI  theory is a modification of the Einstein-Hilbert action which might allow to remove the appearance of singularities, thus avoiding an undesirable feature of Einstein's theory of general relativity (GR). The EiBI theory is expected to be in agreement with GR at energies well below the Planck scale, which represents the regime where quantum gravitational effects are expected to begin to become important and modify the classical description. The naturalness of EiBI gravity has been argued on the basis of canonical procedures to construct Lagrangian densities with second-order field equations \cite{Fiorini}. Moreover, this theory is able to avoid cosmological singularities \cite{Scargill}, has been employed to study properties of dark matter and dark energy \cite{Harko,Banados2,Banados3}, in the coupling to several kinds of fields \cite{Vollick}, as an alternative to inflation \cite{Avelino}, and to explore the structure of compact stars \cite{Delsate,Pani:2012qd}. When coupled to a perfect fluid with a given equation of state, it has been found that the theory can be interpreted as GR coupled again to a perfect fluid, but with a modified equation of state \cite{Delsate2,Pani}.

Though in its determinantal form the EiBI theory may appear as lacking an intuitive motivation, here we show that when applied to elementary systems such as electric fields generated by point-like sources (or elementary particles), the theory boils down to a simple quadratic extension of GR. This simplification occurs when the stress-energy tensor of the matter possesses certain algebraic properties \cite{Stephani}, namely, when it has two double eigenvalues. We take advantage of this property to explore in detail the internal structure of the electrovacuum solutions of the theory and find that the central singularity is generically replaced by a wormhole supported by the electric field. The wormhole structure turns out to be crucial to regularize the total energy stored in the electric field, which occurs in a way that resembles the original Born-Infeld electromagnetic theory. Among the solutions of the theory, there exist a family (characterized by a certain charge-to-mass ratio) for which curvature invariants are finite everywhere. These solutions, whose mass exactly coincides with the energy contained in the electric field, lose the event horizon when the number of charges drops below a critical value, $N_q^c\sim 16$, yielding remnants which are not affected by Hawking's quantum instability. The mass spectrum of these remnants can be lowered from the Planck scale ($\sim 10^{19}$ GeV)  down to the TeV scale if quantum matter corrections are considered. These results are derived in a four-dimensional scenario.

\section{Theory and field equations}

The action of the EiBI theory with matter can be written as
\begin{equation}
S=\frac{1}{\kappa^2 \epsilon} \int d^4x \left[ \sqrt{-\vert g_{\mu\nu} + \epsilon R_{\mu\nu}(\Gamma) \vert } - \lambda \sqrt{-g } \right] + S_m, \label{eq:action}
\end{equation}
where $g$ is the determinant of the space-time metric $g_{\mu\nu}$,  $\vert g_{\mu\nu} + \epsilon R_{\mu\nu}(\Gamma) \vert$ represents the determinant of a rank-two tensor $q_{\mu\nu}\equiv g_{\mu\nu} + \epsilon R_{\mu\nu}(\Gamma) $, $\kappa^2 \equiv 8\pi G/c^4$ is a constant, $\epsilon$ is a (small) parameter with dimensions of length squared,  $R_{\mu\nu}\equiv {R^\rho}_{\mu\rho\nu}$ is the Ricci tensor, ${R^\alpha}_{\beta\mu\nu}=\partial_{\mu}
\Gamma^{\alpha}_{\nu\beta}-\partial_{\nu}
\Gamma^{\alpha}_{\mu\beta}+\Gamma^{\alpha}_{\mu\lambda}\Gamma^{\lambda}_{\nu\beta}-\Gamma^{\alpha}_{\nu\lambda}\Gamma^{\lambda}_{\mu\beta}$ is the Riemann tensor of the connection $ \Gamma_{\mu\nu}^{\lambda}$, which is {\it a priori} independent of the metric (Palatini formalism), and $S_m$ is the matter action. The meaning of the parameter $\lambda$ can be obtained from the field equations and from the leading-order terms of an expansion in $\epsilon \ll 1$, which reads
\begin{eqnarray}
\lim_{\epsilon \rightarrow 0} S &=& \frac{1}{2\kappa^2} \int d^4x \sqrt{-g} \left[R-2\Lambda_{eff} \right] \nonumber \\
&-&\frac{1}{2\kappa^2} \int d^4x \sqrt{-g} \frac{\epsilon}{2} \left(\frac{-R^2}{2} + R_{\mu\nu}R^{\mu\nu}\right)+ S_m \label{eq:GRaction} \ .
\end{eqnarray}
Here $R\equiv g^{\mu\nu}R_{\mu\nu}$ and $\Lambda_{eff}=\frac{\lambda -1}{\epsilon}$. Therefore, when $\epsilon \to 0$ the leading-order in (\ref{eq:GRaction}) coincides with GR plus a cosmological constant term, with $\lambda=1+\epsilon \Lambda_{eff}$. The action  (\ref{eq:action}) can thus be seen as a high-energy modification of Einstein's theory, including a cosmological constant as long as $\lambda\neq 1$.

Variation of (\ref{eq:action}) with respect to metric and connection leads to
\begin{eqnarray}
\frac{\sqrt{\vert q \vert}}{\sqrt{\vert g \vert }} q^{\mu\nu}-\lambda g^{\mu\nu}&=&-\kappa^2 \epsilon T^{\mu\nu} \label{eq:metric} \\
\nabla_{\alpha} \left(\sqrt{q} q^{\mu\nu} \right)&=&0, \label{eq:connection}
\end{eqnarray}
where $q^{\mu\nu}$ is the inverse of $q_{\mu\nu}$. To obtain these equations we have assumed vanishing torsion, $\Gamma_{[\mu\nu]}=0$, as well as $R_{[\mu\nu]}=0$, which guarantees the existence of volume invariants preserved by the theory \cite{or13d}. The connection equation (\ref{eq:connection}) is formally identical to that found in the Palatini version of GR and implies that  $\Gamma_{\mu\nu}^{\lambda}$ can be written as the Levi-Civita connection of $q_{\mu\nu}$, which can be seen as an auxiliary metric tensor associated with the independent connection. Similarly as in other Palatini theories, the relation between $q_{\mu\nu}$ and $g_{\mu\nu}$ is algebraic and given by
\begin{equation}\label{eq:qandg}
\hat{q}=\sqrt{|\hat{\Sigma}|} \hat{\Sigma}^{-1} \hat{g} \ , \ \hat{q}^{-1}=\frac{\hat{g}^{-1}\hat{\Sigma}}{\sqrt{|\hat{\Sigma}|}},
\end{equation}
where we have used a hat to denote matrix representation and have defined
$\sqrt{|\hat{\Sigma}|} \hat{\Sigma}^{-1}\equiv \hat I+\epsilon \hat P$, with $\hat P$ denoting the matrix ${P^\alpha}_\nu\equiv g^{\alpha \mu}R_{\mu\nu}(\Gamma)$. This definition implies $|\hat{\Sigma}|=|\hat I+\epsilon \hat P|$.
Using this notation and Eq.(\ref{eq:metric}), one can easily verify that
\begin{equation}\label{eq:Sigma-T}
\hat{\Sigma}=\lambda \hat{I}-\epsilon \kappa^2 \hat {T} \ ,
\end{equation}
where ${[\hat {T}]_\mu}^\nu\equiv T^{\nu \alpha}g_{\alpha \mu}$. Note that Eq. (\ref{eq:Sigma-T}) allows to obtain the relation between $g_{\mu\nu}$ and $q_{\mu\nu}$ once the matter sources have been specified. The field equations referred to the metric $q_{\mu\nu}$ can be written in a very compact and convenient form noting that $\hat q=\hat g+\epsilon \hat R$ can be written as $\epsilon \hat R \hat{ q}^{-1}=I-\hat g \hat{q}^{-1}$. Using Eq.(\ref{eq:qandg}) we find that  $\hat g \hat{q}^{-1}= \frac{\hat{\Sigma}}{\sqrt{|\hat{\Sigma}|}}$, and taking  Eq.(\ref{eq:Sigma-T}) we finally get
\begin{equation}\label{eq:Rmn-h}
{R_\mu}^\nu (q)=\frac{\kappa^2}{{|\hat\Sigma|}^{1/2}}\left[\LL_G {\delta_\mu}^\nu+{T_\mu}^\nu\right] \ ,
\end{equation}
where ${R_\mu}^\nu (q)= R_{\mu\alpha}(\Gamma)q^{\alpha \nu}$, and we have used the fact that the gravity action in (\ref{eq:action}) can be written as $S_G=\int d^4x \sqrt{-g}\LL_G$ with $\LL_G\equiv ({|\hat\Sigma|}^{1/2}-\lambda)/ (\kappa^2\epsilon)$. It is worth noting that the representation of the field equations given in Eq. (\ref{eq:Rmn-h}) seems to be very general, since it is also valid for other families of Palatini theories with Lagrangians of the form $f(R)/(2\kappa^2)$ and $f(R, R_{\mu\nu}R^{\mu\nu})/(2\kappa^2)$, which include GR as a particular case (with $\hat\Sigma=\hat I$).

From Eq. (\ref{eq:Rmn-h}) one clearly sees that the metric $q_{\mu\nu}$ satisfies a system of second-order differential equations with the matter sources on the right-hand side (recall from (\ref{eq:Sigma-T}) that $\hat\Sigma=\hat\Sigma [\hat{T}]$). Since $q_{\mu\nu}$ is  algebraically related to $g_{\mu\nu}$ through Eq.(\ref{eq:qandg}), it follows that $g_{\mu\nu}$ also satisfies second-order equations. On the other hand, it is easy to see that in vacuum, ${T_\mu}^\nu=0$ and $|\hat\Sigma|=\lambda^4$,   $g_{\mu\nu}$ and $q_{\mu\nu}$ are identical up to a constant conformal factor and that ${R_\mu}^\nu (q)=\frac{(\lambda-1)}{\lambda\epsilon}{\delta_\mu}^\nu$, which is equivalent to
$R_{\mu\nu} (g)=\frac{(\lambda-1)}{\epsilon}g_{\mu\nu}$, thus confirming that $\frac{(\lambda-1)}{\epsilon}$ plays the role of an effective cosmological constant in the full theory. Since the vacuum theory is equivalent to GR with a cosmological constant, no ghost-like instabilities are present in the theory, which is a rather general property of Palatini theories.

\section{Electrovacuum solutions}

We now couple our gravity theory to an electromagnetic field with action
\begin{equation}
S_m=-\frac{1}{16\pi} \int d^4x \sqrt{-g} F_{\mu\nu}F^{\mu\nu}, \label{eq:Maxwell}
\end{equation}
where $F_{\mu\nu}=\partial_{\mu}A_{\nu}-\partial_{\nu}A_{\mu}$ is the field strength tensor. Assuming a spherically symmetric and static electric field, without loss of generality we can choose coordinates such that the line element becomes
\begin{equation}\label{eq:ds2}
ds^2=-A(x)dt^2+\frac{1}{A(x)}dx^2+r^2(x) d\Omega^2 \ .
\end{equation}
It is sometimes useful to use the function $r$ as a radial coordinate, which turns (\ref{eq:ds2}) into $ds^2=-A(r)dt^2+dr^2/B(r)+r^2d\Omega^2$. This replacement is possible as long as the relation between the coordinate $x$ and the radial function $r^2(x)$ is monotonic.  For reasons that will become clear later, we use $x$ instead of $r$ in our analysis. With this choice of coordinates, one can verify that $F^{tx}=q/r^2$ is the only non-zero component of $F^{\mu\nu}$, where $q$ is an integration constant. This leads to ${T_{\mu}}^{\nu}=\frac{q^2}{8\pi r^4} diag(-1,-1,1,1)$,  which inserted in  (\ref{eq:Sigma-T}) yields
\begin{equation} \label{eq:sigma+-}
{\Sigma_{\mu}}^{\nu} = \left(
\begin{array}{cc}
 \sigma^{(\epsilon)}_{+} \hat{I} & \hat{0}  \\
\hat{0} & \sigma^{(\epsilon)}_{-} \hat{I}
\end{array} \right) , \ \sigma^{(\epsilon)}_{\pm}=\lambda \pm \frac{\epsilon \kappa^2 q^2}{8\pi r^4} \ .
\end{equation}
The field equations (\ref{eq:Rmn-h}) then become
\begin{equation}\label{eq:Rmn}
{R_{\mu}}^{\nu}(q)=\frac{1}{\epsilon}\left(
\begin{array}{cc}
\frac{(\sigma^{(\epsilon)}_{-} -1)}{\sigma^{(\epsilon)}_{-} }\hat{I}& \hat{0}  \\
\hat{0}& \frac{(\sigma^{(\epsilon)}_{+} -1)}{\sigma^{(\epsilon)}_{+} }\hat{I} \end{array}
\right)  .
\end{equation}
The strategy now consists on solving for $q_{\mu\nu}$ and then use (\ref{eq:qandg}) to obtain $g_{\mu\nu}$. To proceed, we write a line element for $q_{\mu\nu}$, which according to  (\ref{eq:qandg}) and (\ref{eq:ds2}) takes the form

\begin{equation}
d\tilde{s}^2=-\sigma^{(\epsilon)}_{-} A(x) dt^2+(\sigma^{(\epsilon)}_{-}/A(x))dx^2+\sigma^{(\epsilon)}_{+}r^2(x) d\Omega^2,
\end{equation}
and rewrite it as

\begin{equation}
d\tilde{s}^2=-C(\tilde{r}) e^{2\psi(\tilde{r})} dt^2+1/C(\tilde{r}) d\tilde{r}^2+\tilde{r}^2d\Omega^2,
\end{equation}
where we have defined $\tilde{r}^2=\sigma^{(\epsilon)}_{+}r^2$, $C(\tilde{r})=\sigma^{(\epsilon)}_{-} A(x)$, and $(d\tilde{r}/dx)^2=(\sigma^{(\epsilon)}_{-})^2$. Plugging this ansatz into (\ref{eq:Rmn}), we find that $\psi(\tilde{r})$ is a constant.  With the typical ansatz $C=1-\frac{2G M(\tilde{r})}{c^2\tilde{r}}$ we obtain
\begin{equation}\label{eq:Mr}
M_r\equiv \frac{dM}{dr}=\frac{r^2(\sigma_{+}^{(\epsilon)}-1)\sigma_{-}^{(\epsilon)}}{2\epsilon \sigma_{+}^{(\epsilon)1/2}} \ ,
\end{equation}
where  the relation $d\tilde{r}/dr=\sigma_{-}^{(\epsilon)}/\sqrt{\sigma_{+}^{(\epsilon)}}$, which follows from $\tilde{r}^2=r^2 \sigma_{+}^{(\epsilon)}$, has been used. This relation also allows to find that $(dr/dx)^2=\sigma_{+}^{(\epsilon)}$. We thus conclude that the line element (\ref{eq:ds2}) is completely determined by
\begin{equation} \label{eq:Adxdr}
A(x)=\frac{1}{\sigma^{(\epsilon)}_{-}}\left(1-\frac{2G M(r)}{c^2r \sqrt{\sigma_{+}^{(\epsilon)}}}\right) \ , \ \left(\frac{dr}{dx}\right)^2=\sigma_{+}^{(\epsilon)} \ ,
\end{equation}
with $M(r)$ given by the integration of Eq.(\ref{eq:Mr}). Note that, though this solution is formally equivalent to that obtained by Ba\~nados and Ferreira in \cite{Banados}, in the following sections we shall discuss the physical properties of those solutions with $\epsilon<0$. Therefore the claims of \cite{Banados} regarding the presence of singularities in this theory when $\epsilon>0$ do not apply in our case, as we will explicitly show below. For a more detailed discussion on the influence of signs of the parameters in the theory see \cite{or13b}.

\section{EiBI as quadratic gravity}

To discuss the physics behind the above solutions, we note that Eqs. (\ref{eq:Rmn-h}), (\ref{eq:sigma+-}), (\ref{eq:Mr}), and (\ref{eq:Adxdr}) {\it exactly} coincide with those corresponding to the quadratic Palatini theory
\begin{eqnarray}\label{eq:quadratic}
S&=&\frac{1}{2\kappa^2}\int d^4x \sqrt{-g}\left[R-2\Lambda +a\left(-\frac{R^2}{2}+R_{\mu\nu}R^{\mu\nu}\right)\right] \nonumber \\
&+&S_m
\end{eqnarray}
with the identifications $\epsilon=-2a$ and $\lambda=1+\epsilon \Lambda$ [compare with Eq.(\ref{eq:GRaction})].  The reason for this equivalence lies on the algebraic properties of the action (\ref{eq:action}). Given the linear relation between ${T_\mu}^\nu$ and ${\Sigma_\mu}^\nu$, in a basis in which ${T_\mu}^\nu$ is diagonal the matrix $\hat{P}$ will also be diagonal. If $\hat{P}$ has two double  eigenvalues (in our case  $\hat{P}=diag[p_1,p_1,p_2,p_2]$)  then the fourth order polynomial defined by $|\hat{\Sigma}|=|\hat{I}+\epsilon \hat{P}|$ turns into the second order polynomial appearing in
(\ref{eq:quadratic}) when the squared root is evaluated. Note that this quadratic polynomial coincides exactly with the series expansion of Eq. (\ref{eq:GRaction}), thus implying that all other higher-order terms vanish identically when $\hat{P}$ has this structure.

\begin{figure}[h]
\includegraphics[width=0.5\textwidth]{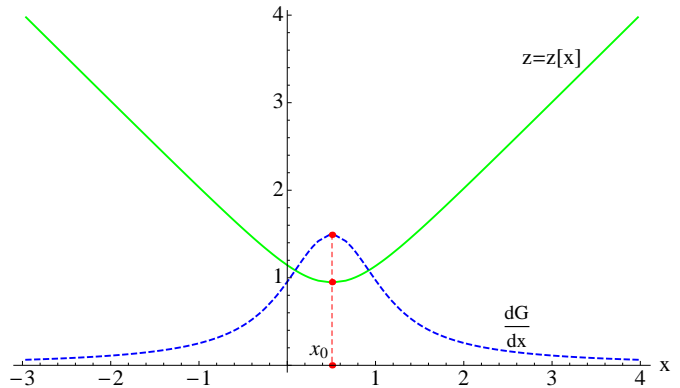}
\caption{Plot of $z(x)$ and $dG/dx$ (with $r_c=1$). The wormhole throat is at the minimum of $z(x)$ (maximum of $G_x$). \label{fig:Bounce}}
\end{figure}

The intimate relation existing between the quadratic Palatini theory and the EiBI theory can be used to shed useful new light on the physics of the corresponding solutions.  To see this, let us focus on the case $\epsilon=-2l_\epsilon^2$, with $l_\epsilon$ some small length scale, for which exact solutions can be found for arbitrary $\lambda$ \cite{PLB13}.  For simplicity, we consider here only the case $\lambda=1$ \cite{or12a,or12d}. The function $M(r)$ appearing in (\ref{eq:Adxdr}) can be written as $M(r)=M_0(1+\delta_1 G(r))$, where $M_0$ is an integration constant representing the Scharzschild mass of the uncharged case, while

\begin{equation} \label{eq:delta1}
\delta_1=\frac{1}{2r_S}\sqrt{r_q^3/l_{\epsilon}}
\end{equation}
is a dimensionless constant that controls the charge-to-mass ratio, and $G(r)$ encodes the electric field contribution. For convenience, we replace $M_0$ and $q$ by the length scales $r_S=2GM_0/c^2$ and $r_q^2=\kappa^2 q^2/4\pi$, and measure the radial function $r(x)$ in units of $r_c\equiv \sqrt{r_q l_\epsilon}$ by defining $z=z(x)=r(x)/r_c$.
For $z\gg 1$, $G(z)\approx -1/z$ leads to the expected GR  limit  $A(x)\approx \left(1-\frac{r_S}{r}+\frac{r_q^2}{2r^2}\right)+O(\frac{r_c^4}{r^4})$ and, therefore, does not exhibit any pathological behavior\footnote{It has been reported recently that stellar models with certain polytropic equations of state may develop curvature divergences at their {\it surface} \cite{Pani:2012qd}, where the interior geometry is matched to an external Schwarzschild metric. Similar problems were already found in the context of Palatini $f(R)$ theories (see \cite{Olmo2008} for a discussion).  We support the view that when curvature divergences arise, a refined (microscopic) description of the troublesome region might help better understand their physical significance. In this sense, the absence of such pathologies in elementary charged systems, as found here, suggests that the results of \cite{Pani:2012qd} might be an artifact of the approximations employed in the continuum description of statistical/macroscopic systems. In fact, since a star is made out of elementary particles, our results indicate that nothing special should happen as the outermost regions are approached, where the effective separation between particles increases and the ``isolated-particle description" of its constituents becomes more and more accurate. The average energy density and gradients in those regions cannot be larger than in the region close to an individual particle  because the volumes involved differ by orders of magnitude. In our view, therefore, a microscopic description of a stellar surface, seen as a collection of elementary particles, seems to be free of the pathologies described in \cite{Pani:2012qd}. As another way out of this problem,  it has been recently argued \cite{Kim} that when the gravitational backreaction on the matter dynamics at the star surface is considered, the effective equation of state gets modified with the consequence that surfaces are no longer singular.}.  For $z\to 1$, however, the geometry strongly departs from the low energy limit represented by GR. To understand the relevance of this region, one should look at the behavior of the function $z(x)=r(x)/r_c$:
\begin{equation}\label{eq:z(x)}
\left(\frac{dz}{dx}\right)^2=\frac{1}{r_c^2}\left(1-\frac{1}{z^4}\right) \ .
\end{equation}
For $z\gg 1$ the relation between $z$ and $x$ is linear, but $z$ reaches a minimum at $z(x_c)=1$. This minimum (or bounce) of the radial function $z(x)$ is a clear signal of the presence of a wormhole. In fact, Eq.(\ref{eq:z(x)}) can be integrated to get the curve shown in Fig. \ref{fig:Bounce}, where $dz/dx=(1-\frac{1}{z^4})^{1/2}/r_c$ if $x\ge x_c$ and $dz/dx=-(1-\frac{1}{z^4})^{1/2}/r_c$ if $x\le x_c$.
Therefore, our space-time can be seen as consisting on two identical pieces connected through a (spherical) hole (the wormhole throat) of area $A=4\pi r_c^2$.
Around $z\approx 1$, $G(z)$ can be expanded as
\begin{equation}\label{eq:G(z)}
G(z)\approx \beta +2\sqrt{z-1}-\frac{11}{6}(z-1)^{3/2}+\ldots
\end{equation}
where $\beta\approx -1.74804$ is an integration constant. With $G(z)$ and $z(x)$ known, the geometry can be explored in detail. One then finds that, in general, there exist curvature divergences at $z=1$, being the leading order dominated by $\sim 1/(z-1)^3$ in the case of  $R_{\mu\nu}(g)R^{\mu\nu}(g)$ and ${R^\alpha}_{\beta\mu\nu}(g){R_\alpha}^{\beta\mu\nu}(g)$ (which contrasts with the much stronger divergences found in GR, $\sim 1/r^8$). However, if the charge-to-mass ratio defined by the constant $\delta_1$ is tuned to the value $\delta_1^*=-1/\beta$, the $z=1$ divergences disappear yielding a completely regular geometry.

\section{Charge without charges and mass without masses}

One can now wonder about the nature of the sources that generate the charge $q$ and the mass $M_0$ that characterize our solutions\footnote{Note that in the context of GR it is sometimes stated that the charge and mass are concentrated at a point of zero volume at the center (the singularity). However, the fact is that there is no mathematically well-defined source able to generate the Reissner-Nordstr\"{o}m solution \cite{ortin}.}. As first shown by Misner and Wheeler \cite{Misner-Wheeler}, an electric flux through a wormhole can define by itself an electric charge\footnote{This charge is a very primitive concept that does not require for its existence neither the definition of  metric nor affine structures on the manifold and, as such, is insensitive to the presence of curvature divergences.} without the need for sources of the electric field. Therefore, the charge $q$ appearing in our solutions is entirely given by the electric flux through any two-surface $\mathcal{S}$ enclosing one of the sides of the wormhole, i.e.,

\begin{equation}
q=\frac{1}{4\pi}\int_{\mathcal{S}} *F,
\end{equation}
where $*F$ is the two-form dual to Faraday's tensor. This charge is conserved as long as the topology does not change.

The existence of a wormhole where one would naively expect to find the sources poses a more severe challenge to identify the origin of the mass $M_0$. A tempting guess would be to see $M_0$ as related to the energy stored in the electric field. This idea, however, seems in conflict with the results from Minkowski space-time, where the energy of a point-like charged field is defined as $\E_e= 4\pi \int_0^{\infty}dr r^2 q^2/8\pi r^4 $, and diverges as $r\to 0$. Historically, this divergence was circumvented by replacing Maxwell electrodynamics (\ref{eq:Maxwell}) by Born-Infeld theory \cite{BI}, whose Lagrangian reads, in determinantal form
\begin{equation}
\LL_{BI}= \beta^2\left(\sqrt{-\vert g_{\mu\nu} + \beta^{-1} F_{\mu\nu} \vert} - \sqrt{-g}  \right) \label{eq:BI}
\end{equation}
This modification yields a total finite energy for the electromagnetic field,

\begin{equation}
E_e^{BI}=n_{BI}  q^{3/2}(c\beta^2)^{1/4},
\end{equation}
where $n_{BI}=\frac{\pi^{3/2} }{3\Gamma(3/4)^2}\approx 1.23605$. Given that in our gravitational model the function $r$ is bounded to $r\ge r_c$, the electric energy could be somehow regularized by the non-trivial topological structure of the space-time. Since for an electric field in Minkowski space the action can be written as $S_{Maxwell}= \int dt \times \E_e$,  to estimate the total electric energy in a gravitational scenario we propose to evaluate the total action as a means to get $S=\int dt \times (\E_G+\E_e)$. By doing this, we find that

\begin{equation}
S=\frac{4\pi r_c^3}{l_\epsilon^2 \kappa^2}\alpha\int dt,
\end{equation}
where $\alpha=\int_1^\infty dz \frac{z^4+1}{z^4\sqrt{z^4-1}}$. Evaluating this integral, we find that $\alpha=\frac{\sqrt{2}\pi^{3/2} }{3\Gamma(3/4)^2}\approx 1.74804=1/\delta_1^*=\sqrt{2} n_{BI}$. With simple manipulations, one finds that the total energy can be written as $\E_T\equiv (\E_G+\E_e)= 2M_0c^2 \delta_1/\delta_1^*$, where the factor $2$ stems from the need to integrate on both sides of the wormhole. Remarkably, this result is finite regardless of the value of $\delta_1$, which implies that the total energy is insensitive to the presence of curvature divergences. We note that these objects, with their charge having a topological origin and their mass being generated by the electric field, naturally realize the idea of {\it geon} (self-consistent solutions of the sourceless  gravito-electromagnetic field equations) originally introduced by Wheeler \cite{Wheeler}.

\section{Horizons and remnants}

When $\delta_1=\delta_1^*$, from Eq.(\ref{eq:delta1}), it is easily seen that the mass spectrum can be written as
\begin{equation}\label{eq:massspectrum}
M_0=n_{BI}\left(\frac{N_q}{N_q^c}\right)^{3/2}m_P \left(\frac{l_P}{l_\epsilon}\right)^{1/2} \ ,
\end{equation}
where $N_q \equiv q/e$ represents the number of charges (with $e$ being the electron charge), $N_q^c=\sqrt{2/\alpha_{e.m.}}\approx 16.55$  (with $\alpha_{e.m.}$ the fine structure constant) is the \emph{critical number of charges} and plays an important role in the existence or not of event horizons [see paragraph below], $m_P$ is the Planck mass, and $l_P$ is the Planck length. It is worth noting that defining the length scale $l_\beta^2\equiv (4\pi/\kappa^2 c\beta^2)$, we can write the expression for $\E_{BI}$ given above as

\begin{equation}
\frac{E_{BI}}{c^2}=\sqrt{2}n_{BI}\left(\frac{N_q}{N_q^c}\right)^{3/2}m_P \left(\frac{l_P}{l_\beta}\right)^{1/2},
\end{equation}
which, up to a factor $\sqrt{2}$, is identical to (\ref{eq:massspectrum}) with the replacement of the electromagnetic scale $l_\beta$ by the gravitational scale $l_\epsilon$. Remarkably,  (\ref{eq:massspectrum}) was derived using the Maxwell electromagnetic Lagrangian. Thus, besides regularizing the geometry, the EiBI gravitational theory  also regularizes the matter sector in a way almost identical to its electromagnetic counterpart.

A very important aspect of the $\delta_1=\delta_1^*$ solutions is that, as can be graphically verified (see \cite{or12d} for details), for $N_q\leq N_q^c$ there is no event horizon, which implies stability of that sector of the theory against Hawking radiation. For $N_q>N_q^c$ the horizon exists and its location almost coincides with the prediction of GR. This means that black holes, understood as objects with an event horizon, can be continuously connected with horizonless configurations lying in the lowest part of the charge and mass spectrum. Such states can be naturally identified as black hole remnants and their existence could have deep theoretical implications for the information loss problem and the Hawking evaporation process \cite{Fabbri}. Note, in addition, that for these remnants the surface $z=1$ is timelike and $S=2M_0c^2\int dt$ coincides with the action of a point-like particle at rest, which suggests that they possess particle-like properties.

\section{Coupling of BI matter}

In order to test the robustness of the results obtained so far against quantum corrections in the matter sector, one can work within the effective Lagrangians approach and consider the coupling of the EiBI gravity model to some nonlinear theory of electrodynamics. The Born-Infeld electromagnetic Lagrangian (\ref{eq:BI}) is a well-motivated choice which, in turn, allows to find exact analytical solutions \cite{or13d}. One then finds that the global qualitative picture provided by Maxwell's theory is preserved but with relevant quantitative differences. In particular, following the same procedure as in the Maxwell case, the mass spectrum (\ref{eq:massspectrum}) turns now into
\begin{equation}
M_0^{BI}=\left(\frac{4\gamma}{1+4\gamma}\right)^{1/4}M_0,
\end{equation}
where $\gamma=(l_\epsilon/l_\beta)^2$.
If the quantum effects of the matter manifest themselves much earlier than the quantum effects of gravity, i.e, if $l_\beta\gg l_\epsilon$, then $M_0^{BI}$ may  become much smaller than $M_0$. In particular, if $l_\epsilon=l_P$ and $\beta$ is pushed to the limits of validity of the effective Lagrangians scheme of quantum electrodynamics, $\beta\sim 10^{18}$, then $\gamma\sim (10^{-17})^4$ and the mass spectrum for $M_0^{BI}$ becomes $17$ orders of magnitude smaller than the original $M_0$. This means that the mass spectrum may drop from $M_0\sim m_P\sim 10^{19}$ GeV  down to $M_0^{BI}\sim 10^2 $ GeV. Though more accurate descriptions of the matter sector might alter these numbers, the fact is that new quantum gravity phenomenology within the reach of current particle accelerators arises within a purely four-dimensional scenario.

\section{Conclusions and outlook}

We have shown that for spherically symmetric charged systems EiBI theory recovers the GR predictions for $r\gg r_c$. The theory, however, changes the microstructure of the space-time replacing the GR singularity by a wormhole. Though curvature divergences may exist at $r=r_c$, their role is uncertain, since they affect neither the properties of the flux through the wormhole nor the finiteness of the total electric energy. The theory makes definite predictions about the existence of black hole remnants and their mass spectrum, with nontrivial implications for the Hawking evaporation process, the information loss problem, and potentially new dark matter candidates \cite{lor}.

\section*{Acknowledgments}

Work supported by project FIS2011-29813-C02-02 (Spain), the Consolider Program CPANPHY-1205388, the JAE-doc program and i-LINK0780 project of CSIC, the CNPq projects 561069/2010-7 and 301137/2014-5 (Brazil), and the Erwin Schr\"odinger fellowship Nr. J3392-N20 of FWF (Austria).

\end{document}